\documentclass{ws-rv961x669}
\usepackage{ws-rv-van}     
\usepackage{ws-rv-thm}     
\makeindex

\begin{document}

\chapter[]{Personal History with MEF and Some Related Topics}

\author[]{Helen Au-Yang\footnote{Helen's part of the story is told by her coauthor and husband of 40 years, as, at the time of writing, she was not (yet) sufficiently recovered from a stroke.} and Jacques H.H. Perk}

\address{Department of Physics, Oklahoma State University,\\ 
145 Physical Sciences, Stillwater, OK, 74078-3072, USA \\
perk@okstate.edu}

\begin{abstract}
We present our personal histories with Michael Fisher. We describe how each
one of us first came to Cornell University. We also discuss our many subsequent
interactions and successful collaborations with him on various physics projects.
\end{abstract}


\body



\section{From Shanghai to Cornell University}\label{sec1}

Born under the name Ou Yang Yee Sun in Shanghai and after having suffered through much of the great famine of 1959--1962, Helen got official permission to join her parents in Hong Kong. It was quite a shock, not knowing Cantonese and English, as she only spoke Shanghainese, Mandarin and some Russian that she had learned in high school. Fortunately, the written Chinese is universal and Helen found a poster announcing a competition for two fellowships at Chu Hai College, Kowloon. Even though Helen had not yet finished high school, she won one of them and was admitted under the new name Helen Au-Yang. After graduating in 1965 she was admitted to San Diego State University.

In 1968 Helen entered graduate school at the State University of New York in Stony Brook. Here she generalized the work of McCoy and Wu\cite{McCoyWu67} on the Ising model on the half plane with boundary magnetic field to the case with boundary coupling different from the coupling in the bulk. Thus she provided exact results\cite{Auyang73} related to the series, Monte Carlo and mean field results of Binder and Hohenberg\cite{BinderHohenberg72,BinderHohenberg74} on 2D and 3D Ising and Heisenberg models with a boundary.

Helen continued with the main topic of her PhD thesis\cite{Auyang74}: Layered Ising models with period $n$, deriving exact results for the specific heat\cite{Auyang74a} and for the pair correlation in a row parallel to the layering\cite{Auyang74b}. That last calculation led to a block Toeplitz determinant with a $2\times2$ matrix generating function $a(\xi)$. After reviewing the then known theorems for such determinants, especially for those for which $a(\xi)$ can be properly factorized, Helen gave $n=2$ results for the $T<T_c$ spontaneous magnetization and for the leading long-distance behavior of the pair correlation for $T<T_c$ and for $T>T_c$.

In 1973 Helen graduated with a PhD in Theoretical Physics, with her thesis signed by Professors B. M. McCoy, C. N. Yang, T. T. Wu and others.\footnote{The publication of the thesis\cite{Auyang74} and the papers\cite{Auyang74a,Auyang74b} got delayed, as her advisor Barry McCoy was involved with the massive project on Painlev\'e III scaling functions in 2D Ising correlations.\cite{WuMcCoyTracyBarouch76}} She became a postdoc with Michael Fisher at Cornell University.


\section{Helen at Cornell University}

With Helen in his group, Michael Fisher continued the work on his project
``Bounded and inhomogeneous Ising models" started with Arthur Ferdinand\cite{FisherFerdinand67,FerdinandFisher69} many years earlier.
In 1969 they had published finite-size effects on the specific heat of 
an $m\times n$ Ising lattice, asymptotically for large $n$ with $\xi=m/n$ fixed.\cite{FerdinandFisher69}

Helen and Michael first studied the specific-heat scaling function for an infinitely long Ising strip of width $n$ in the limit $n\to\infty$.\cite{AuyangFisher75}
 After this, they studied the finite-size effects of regularly spaced point defects,\cite{FisherAuyang75,AuyangFisherFerdinand76,Auyang76} also incorporating some earlier calculations of Ferdinand.

Michael was very happy to have Helen around, as he liked things to be done exactly, if they could be done exactly. One time a student solved a cubic equation numerically by computer. Michael told Helen,``Helen, you show him how that is done exactly!"

Michael wanted to keep Helen for a longer time by creating a special research position for her. He asked C.~N.~Yang to take her back as a postdoc for a year, while he created such a new position at Cornell. At Stony Brook Helen calculated Ising multispin correlations along the diagonal\cite{Auyang77a} to check the operator reduction formulae of Kadanoff. She also calculated a new scaling form for the four-spin correlation along the diagonal.\cite{Auyang77b}


Back at Cornell University, Helen got involved in a generalization of earlier work,\cite{AuyangFisher75} namely the study of an $n\times\infty$ Ising strip with field $h_1$ on the first infinite layer and field $h_n$ on the last ($n$th) layer. One motivation for this work was to investigate the validity of the \textit{ad hoc} local free energy functional of Fisher and de Gennes\cite{FisherDegennes78}. In spite of serious limitations of that theory, their most striking predictions were confirmed and finite-size scaling forms were derived in the limit $n\to\infty$.\cite{AuyangFisher80,FisherAuyang80} A more detailed summary of this research is given in section 4.3 of Fisher's ``Simple Ising Models Still Thrive!'' lecture.\cite{Fisher81}

At the same time period much numerical work was done to study inhomogeneous differential approximants for power series of one or more variables\cite{FisherAuyang79} Nowadays, with Maple and Mathematica available, this is a lot easier, but then it was not. At that time using Fortran, one had to prepare a stack of punchcards for each example with up to 80 characters per card and hand that in for processing overnight, carefully keeping the stacks of cards in the right order. This work was very labor intensive, as not only existing high- and low-temperature series coefficients and activity (high-field) expansions of various well-known spin models were examined, but also many test examples derived from mathematical functions with a given singularity and various background noise contributions.

One case was studied in detail: Universality tests at Heisenberg bicritical points.\cite{FisherChenAuyang80} The experimentally determined amplitude ratio
$Q_{\mathrm{fit}}\approx 1.6\pm0.35$ found in MnF$_2$ was much smaller than theory
seemed to predict, $Q_{\mathrm{th}}\approx 2.39$. Noise contributions in the theory
could bring that value about $10\%$ down, but that is not enough. Fisher et al.\ then
note\cite{FisherChenAuyang80} that shifts in the critical lines compatible with
experimental uncertainties could raise $Q_{\mathrm{fit}}$ enough to bring theory
and experiment in agreement. In other words, $Q_{\mathrm{fit}}$ should have been
reported with significantly larger error bars.


During a visit back to Stony Brook, Helen calculated the scaling form for the four-spin correlation function of two parallel nonaligned pairs in the planar Ising model.\cite{Auyang81}

Some time in 1980, Helen's oldest sister and her family were released from labor camp in Western China to settle in Hongkong with the rest of the family there. Helen then abruptly quit her position at Cornell in order to help tutor her two nieces into the Hongkong school system, using the earlier experience that she had had moving from Shanghai to Hongkong in 1962. In 1982 Helen wanted to return to the USA, and Barry McCoy and Joel Lebowitz organized a special invitation to the December 1982 Rutgers Stat.~Mech.~Meeting.


\section{Jacques' visits to Cornell and Rutgers}

In 1978 I (Jacques) received the Royal Dutch Shell Oil Travel Prize awarded at Leiden University for my thesis research. This award allowed me to visit research groups at several universities and research institutes in the USA. Thus I came to also visit Professors Fisher and Widom at Cornell University, giving a seminar on (in)stability of critical behavior,
\cite{CapelPerkDenouden78,CapelDenoudenPerk79} as that seemed more suitable than my thesis work on time-dependent correlations in XY chains. Students and postdocs had beforehand told me that Fisher had ``destroyed" the speakers in the seminar the previous several weeks, but that did not happen this time. I had seen Fisher in action at earlier conferences and I was prepared. I got a few questions about how things were proved that I could answer.

Noteworthy was the party at Fisher's home. It started with a box of Chinese metal wire puzzles put in front of me. As I had about the same collection, I knew these well. I quickly took a few apart and put them together again, upon which Fisher removed the box, causing a sigh of relief among the others. Fisher then got his Spanish guitar out. He was quite accomplished and entertaining. This
 and the special treats Mrs.~Fisher had prepared were the high points of the party.
 
At the December 1979 Stat.~Mech.~meeting at Rutgers, I asked Leo Kadanoff if he knew that there are continuously varying critical exponents if one varies the coupling constants within one line in the bulk of the square-lattice Ising model.\cite{Bariev79,McCoyPerk80,McCoyPerk82} With Michael Fisher and others standing by, Kadanoff said that that is impossible, upon which I said to expect a preprint soon. Fisher and Kadanoff followed that up with some further remarks on the phenomenon. \cite{Fisher81a,Kadanoff81}

At this point, we note that, like many others, both of us were very surprised that Fisher and Kadanoff did not share the 1982 Nobel Prize in Physics with Kenneth Wilson.


\section{Helen and Jacques married}

It all started with Barry McCoy asking Jacques to pick up Helen, who had visited Michael Fisher at Cornell, at the Greyhound bus terminal in Manhattan on New Year's eve of 1982 and to bring her to the McCoy home. Jacques and Helen did not know that Barry and Martha McCoy conspired and had a detailed plan designed to get both of us married. All the scheming worked and six days later we were engaged to be married, with the wedding taking place on January 22, beginning a very happy marriage lasting more than 40 years so far.

Soon after Helen asked, ``Jacques, have you ever done anything with your quadratic difference equations for 2D Ising correlations?" This led to the first of many joint papers following, namely a letter with fairly complete results to calculate critical Ising pair correlations\cite{AuyangPerk84} and even a result for the monomer-monomer correlation in the square dimer problem. Many years later, students in the statistical mechanics class of a colleague told us that this letter was cited in Pathria's textbook.\cite{PathriaBeale11}
 

Having seen several papers on chiral clock models written at Cornell University during her time there, Helen was led to extend the search for integrable Potts models to chiral ones, discovering the genus 10 curve condition for solving the star-triangle equations for the 3-state non-selfdual chiral Potts model.\cite{AuyangMcCoyPerkTangYan87} The full parametrization of the $N$-state case was found during a visit to Canberra\cite{BaxterPerkAuyang88} and the history of these discoveries is described in a Topical Review celebrating Baxter's 75th birthday and implicitly Helen's 70th.\cite{Perk16}


\section{Later collaborations with Michael Fisher}

Michael Fisher had been very apprehensive about our marriage at the beginning, but he noted that Helen was very happy, seeing her at several of Lebowitz's Stat.~Mech.~meetings at Rutgers.

\begin{figure}[ht]
\centerline{\includegraphics[width=13 cm]{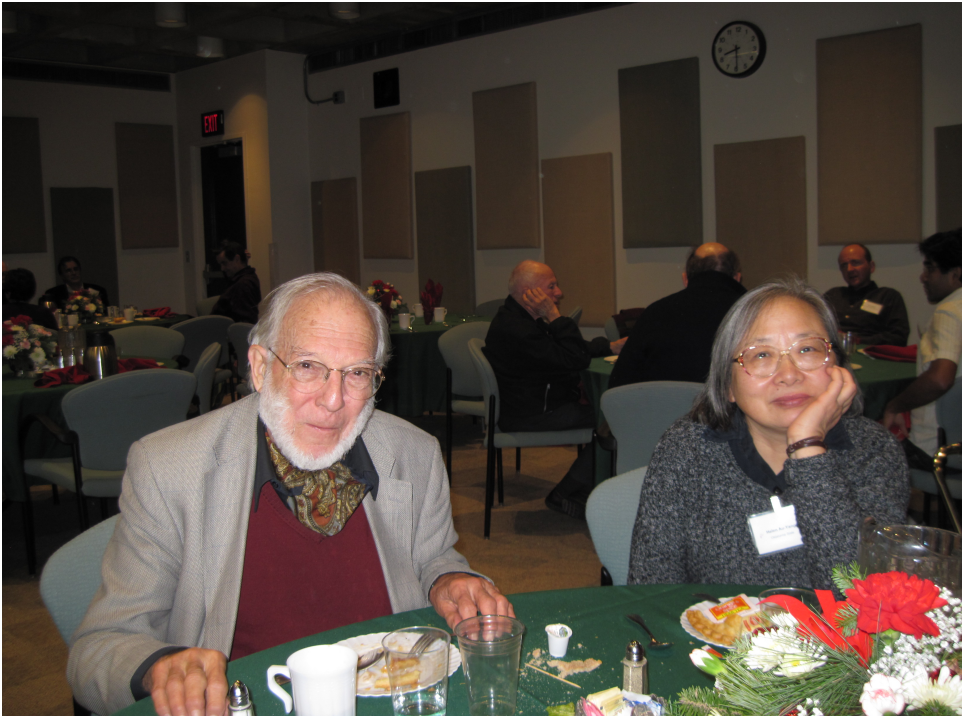}}
\caption{Helen and Michael at December 2013 Stat.~Mech.~Meeting at Rutgers University.}
\label{fig0}
\end{figure}

Then, at the end of 2008, Fisher called me (Jacques) that he wanted a calculation done and that Helen was the only one he trusted to do it correctly. I told him that he was free to ask Helen and let her decide what to do. Fisher then emailed a letter in his typical style, which is reproduced in the appendix. Fisher wanted some exact calculations done on the specific heat to qualitatively understand some features seen in the experiments of Gasparini's group in Buffalo.\cite{GaspariniKimballMooneyDiaz-Avila08,PerronGasparini12}

There was some delay, but in the end two papers were published on 2D Ising models with alternating layers with weaker and stronger interactions,\cite{AuyangFisher13,Auyang13}
the second with the details left out in the first one. Fisher was very happy that I helped with some of the LaTex coding and cleaning up the figures.

Later, Helen thought of a better way to compare with an array of 3D cubes of Helium covered with a 2D film. Correspondingly she studied 2D Ising layers connected by layers of 1D Ising strings, with the strings 1, 2, or 3 units apart. First she did a 60 page calculation of the free energy in her usual fine print handwriting using Hamm's method\cite{Hamm77}, followed by a 20 page calculation, reproducing the results using another method\cite{AuyangFisherFerdinand76}. When she showed me the results, we could guess the general answer for Ising strings $N$ units apart and next prove it\cite{AuyangPerk18a,AuyangPerk18b}. We also obtained exact results for the spontaneous magnetization.\cite{AuyangPerk18a,AuyangPerk19}

Finally, Fisher contacted me, as he had to do something about the 3D Ising deceptions of Zhang Zhidong. He asked me if I had any follow-up on my original comment. I sent him what I had, including my last comment.\cite{Perk13} Fisher got also very upset that his friend of long ago, Norman March, was fooled by the deceptions. Together we wrote another comment\cite{FisherPerk16}, in which we also introduced the statistical mechanics community to the new bootstrap results for 3D Ising critical exponents. Fisher was particularly happy, when I pointed out that the result for the correction to scaling exponent $\Delta=\omega\nu=0.5231(12)$ is very close to the $\frac12$ proposed by Andrea Liu and him\cite{LiuFisher90}

In conclusion, it has been a great pleasure to work with Michael Fisher. In some sense Michael was the conscience of the statistical mechanics community, keeping people honest with probing questions during and after their talks and often making good suggestions. He will be dearly missed.


\newpage
\section*{Appendix}

\begin{figure}[ht]
\centerline{\includegraphics[width=12cm]{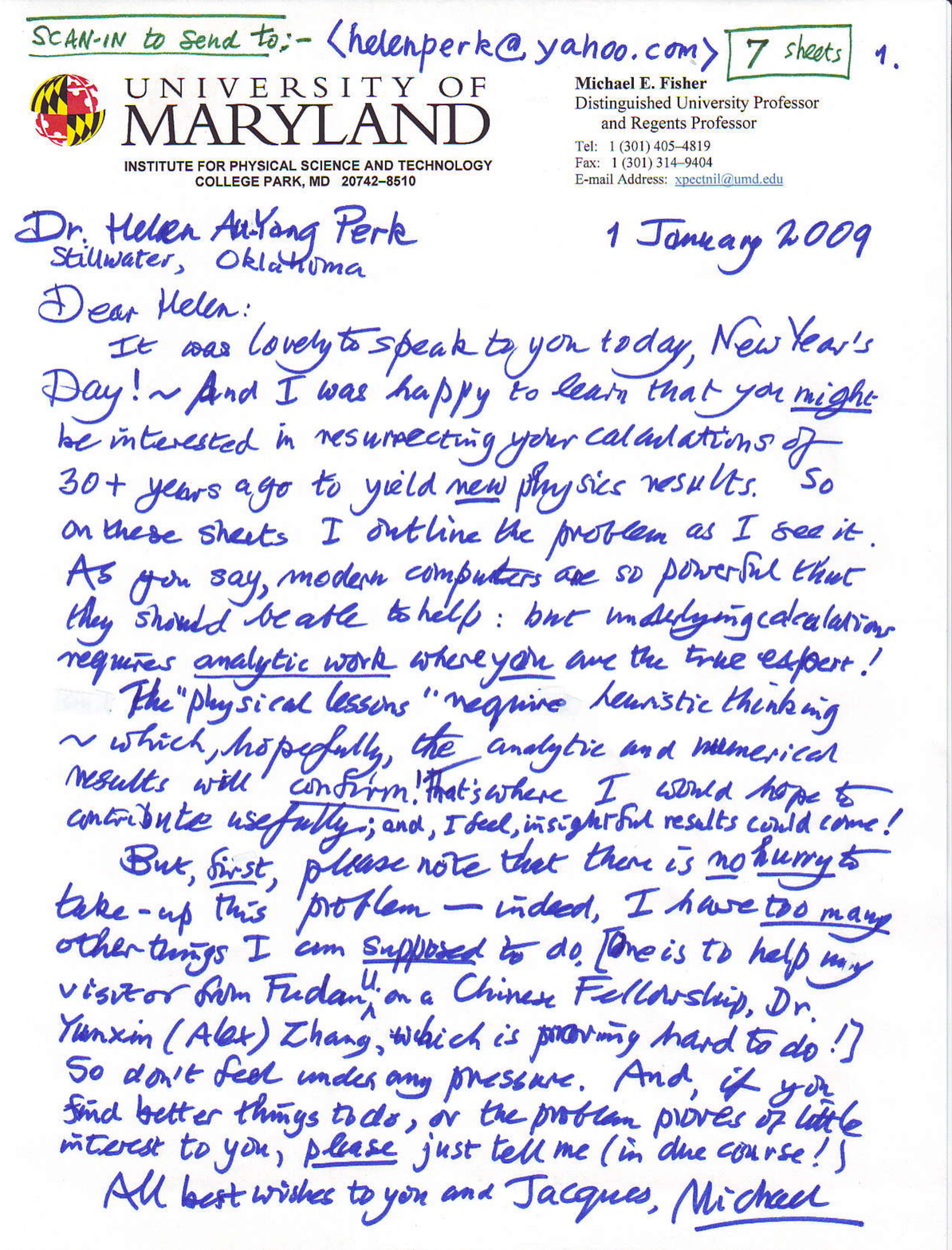}}
\caption{Page 1 of Fisher's January 2009 email to Helen.}
\label{fig1}
\end{figure}

\begin{figure}[ht]
\centerline{\includegraphics[width=13cm]{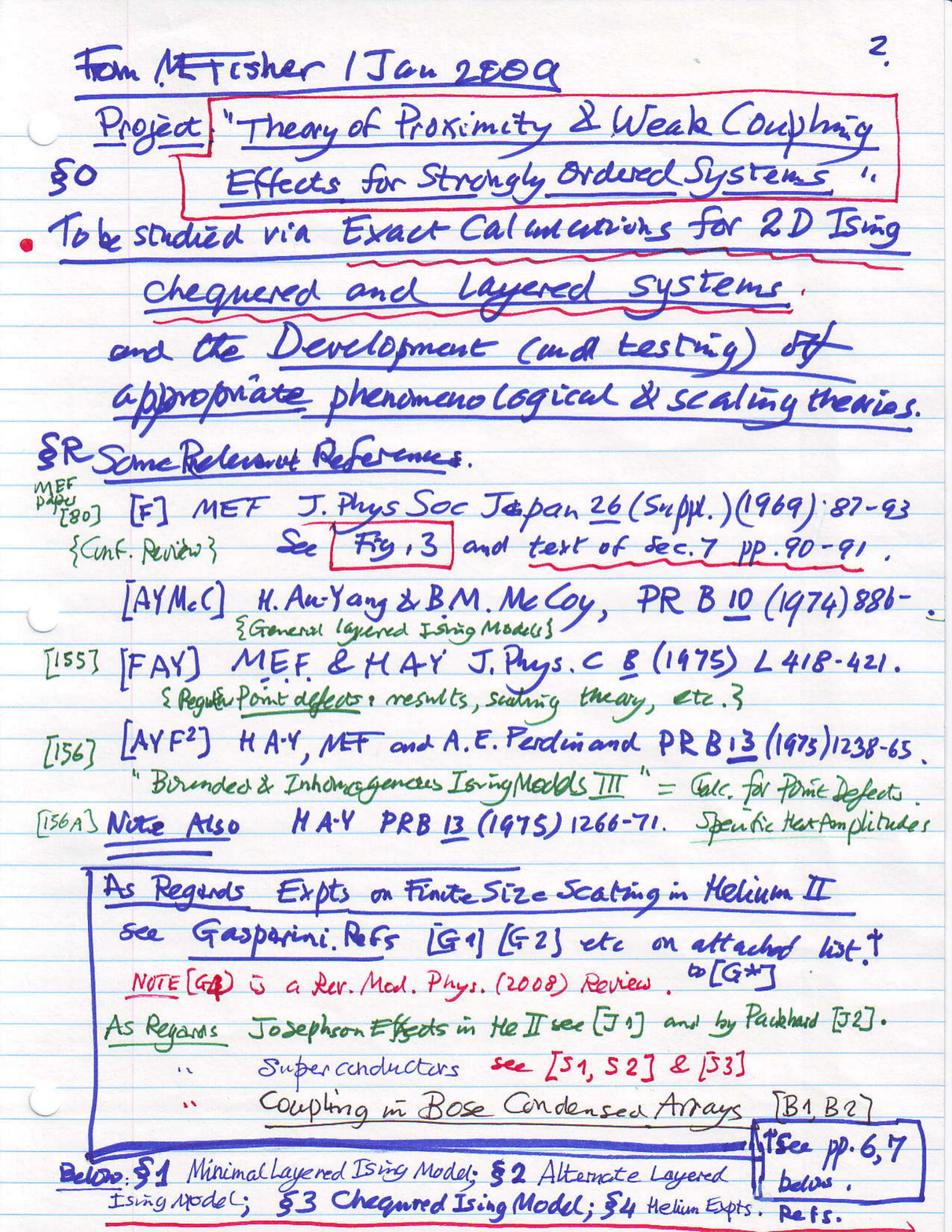}}
\caption{Page 2 of Fisher's January 2009 email to Helen.}
\label{fig2}
\end{figure}

\begin{figure}[ht]
\centerline{\includegraphics[width=13cm]{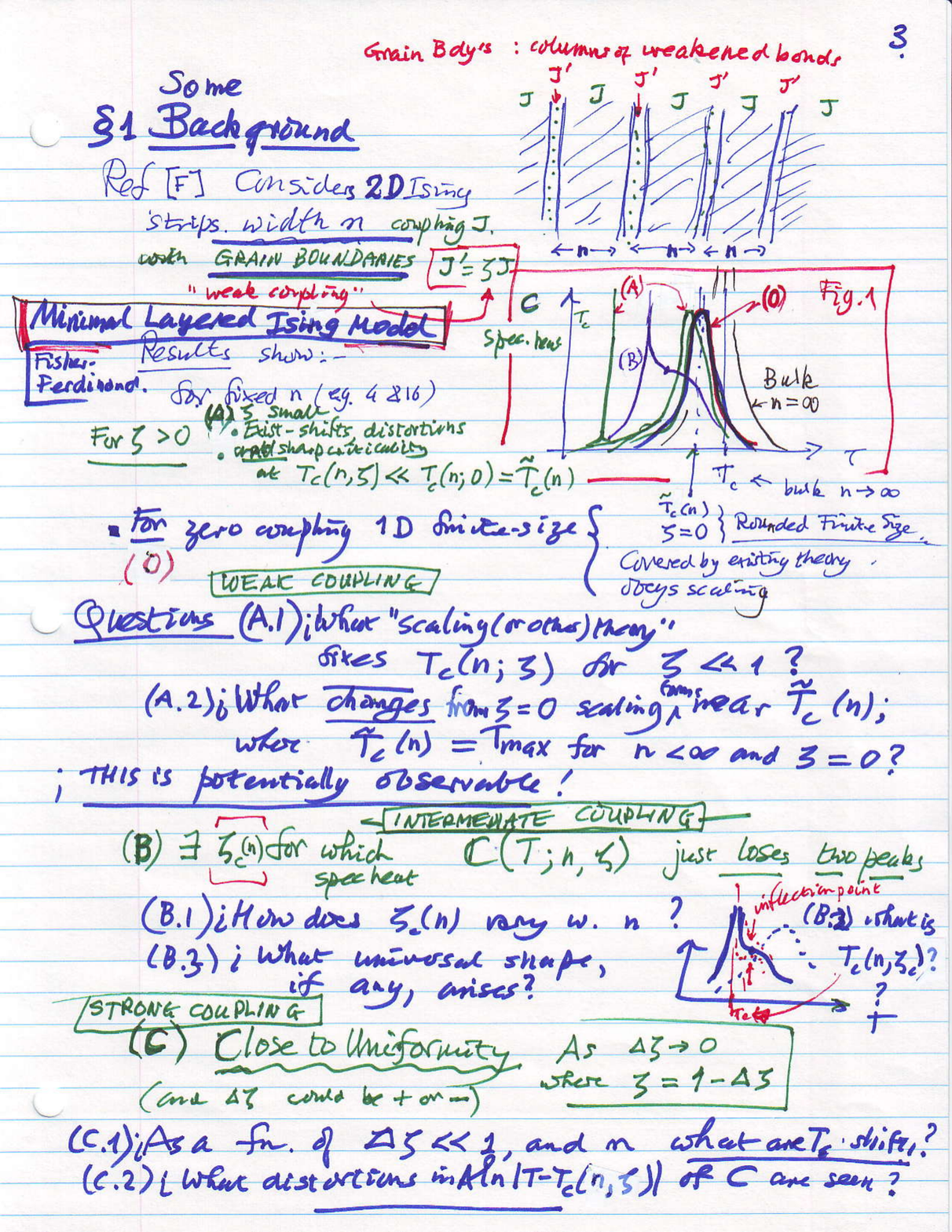}}
\caption{Page 3 of Fisher's January 2009 email to Helen.}
\label{fig3}
\end{figure}

\begin{figure}[ht]
\centerline{\includegraphics[width=13cm]{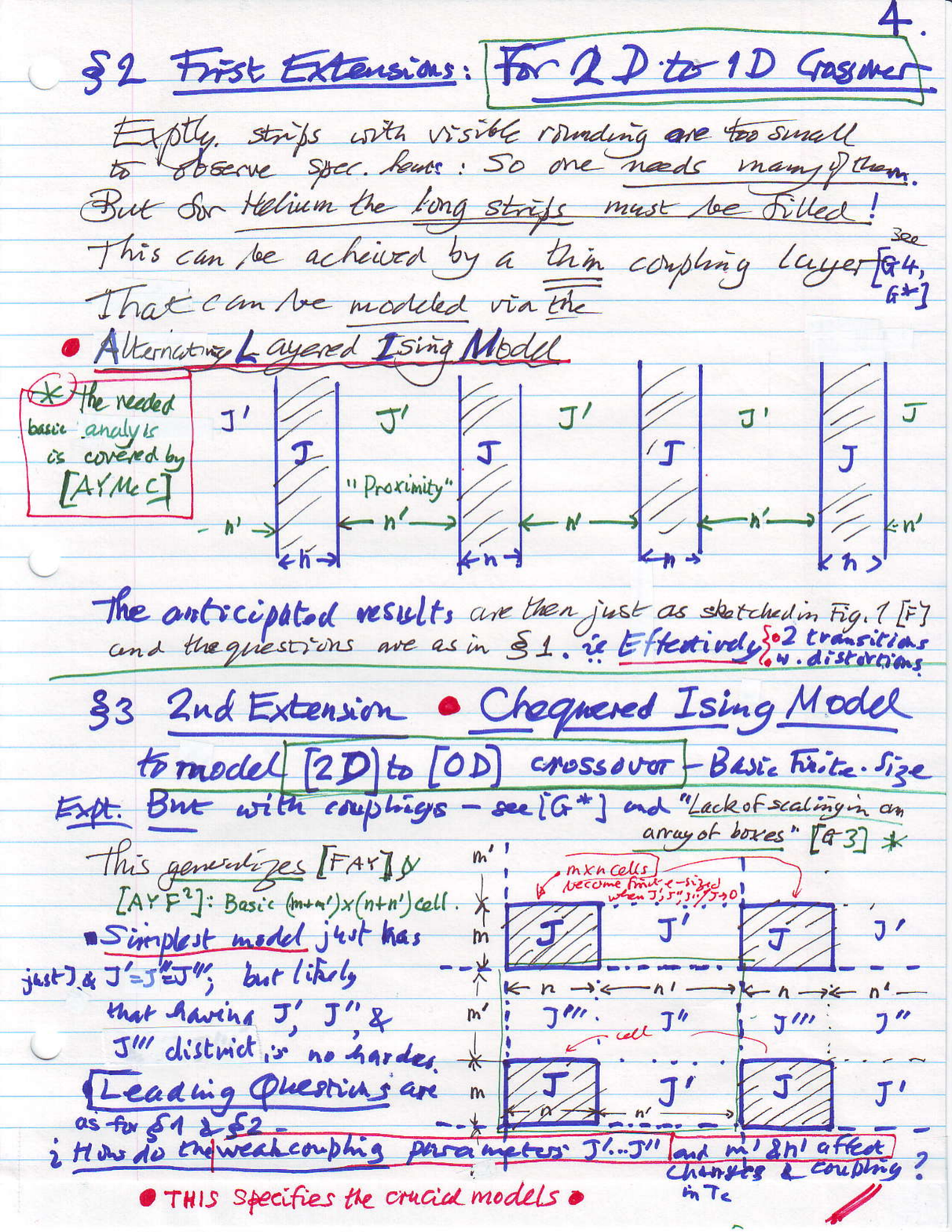}}
\caption{Page 4 of Fisher's January 2009 email to Helen.}
\label{fig4}
\end{figure}

\begin{figure}[ht]
\centerline{\includegraphics[width=13cm]{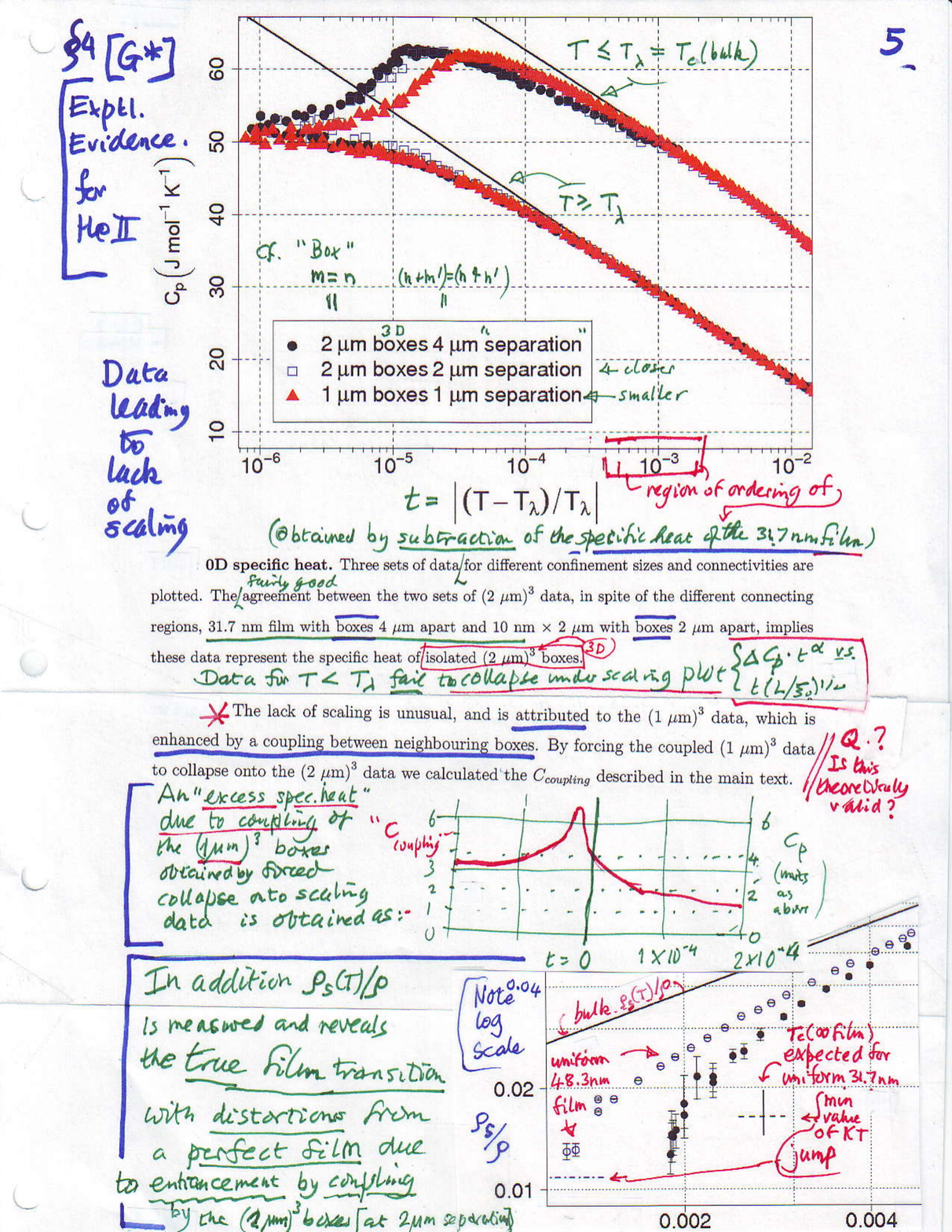}}
\caption{Page 5 of Fisher's January 2009 email to Helen.}
\label{fig5}
\end{figure}

\begin{figure}[ht]
\centerline{\includegraphics[width=13cm]{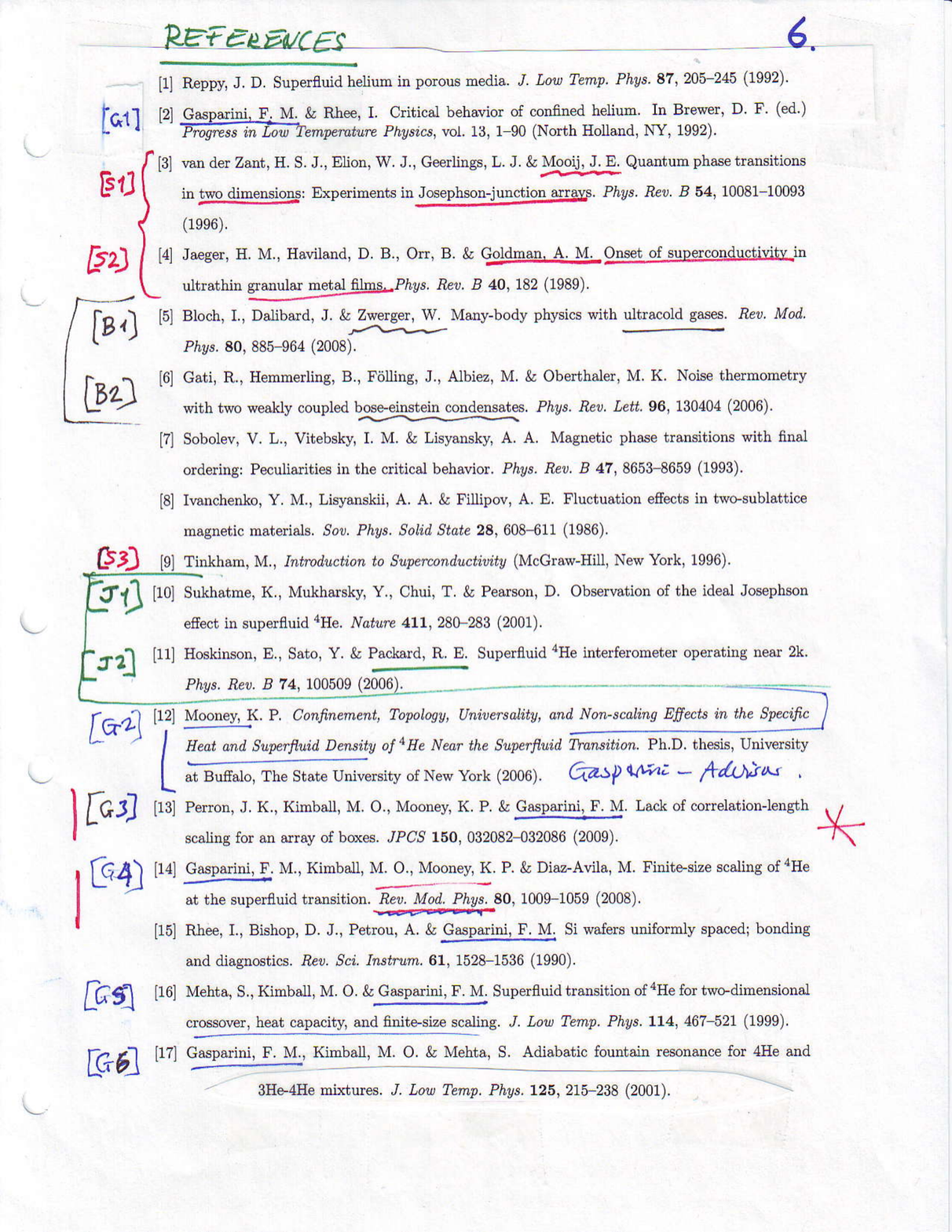}}
\caption{Page 6 of Fisher's January 2009 email to Helen.}
\label{fig6}
\end{figure}

\begin{figure}[ht]
\centerline{\includegraphics[width=13cm]{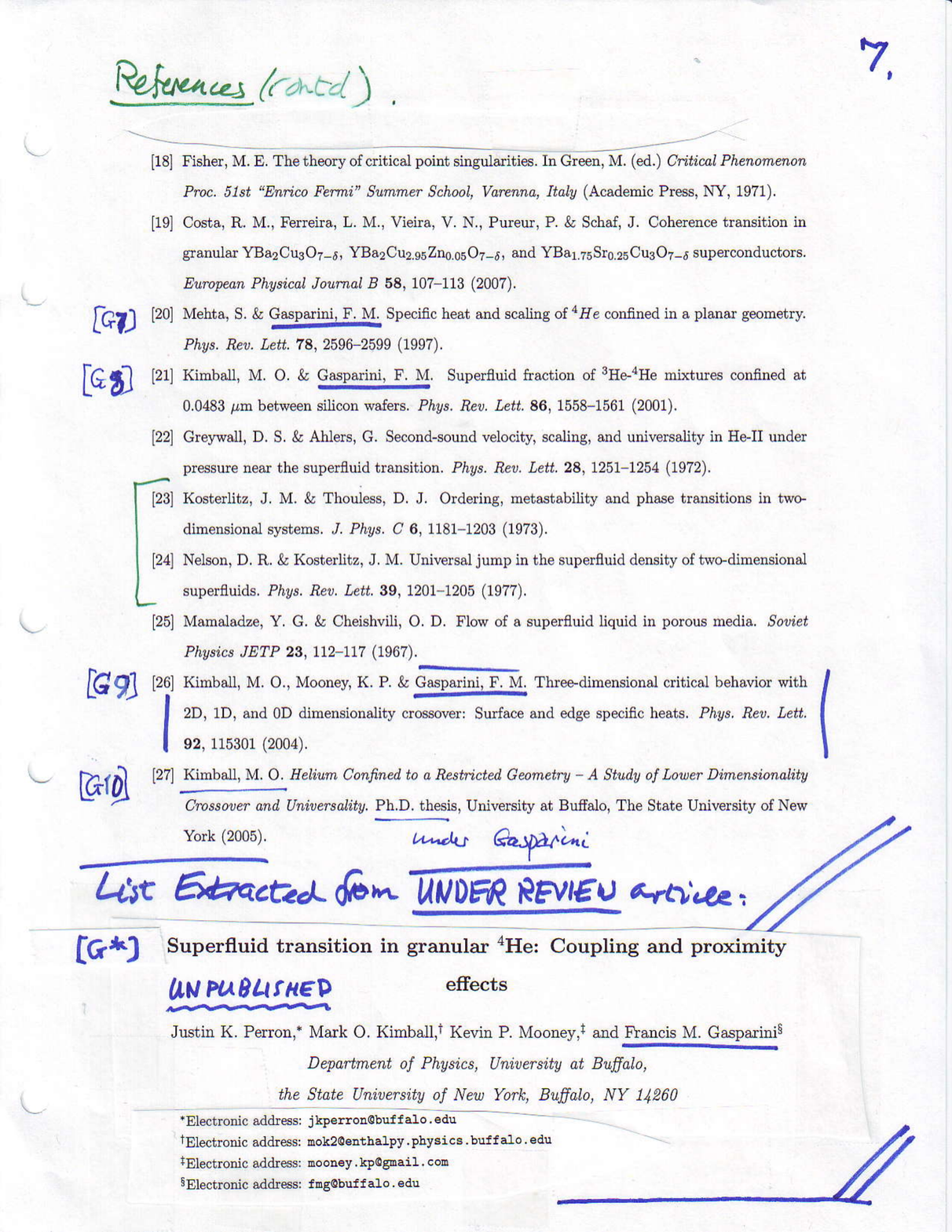}}
\caption{Page 7 of Fisher's January 2009 email to Helen.}
\label{fig7}
\end{figure}

\end{document}